# Gate-controlled photo-oxidation of graphene for electronic structure modification†

Ryo Nouchi,*[a,b] Morihiro Matsumoto [a] and Nobuhiko Mitoma [a]

Graphene is an ultrathin material, which allows us to control surface phenomena by means of field-effect gating. Among various surface phenomena, photo-oxidation is known to be a facile method to largely control the electronic structure of graphene. In this study, gate controllability of photo-oxidation of graphene is thoroughly examined using a field-effect-transistor configuration. The presence of water molecules enhances gate controllability, which can be explained using water-oxygen co-adsorption picture. In addition, the photo-oxidation reaction evolves from the edge and proceeds towards the center of the graphene channel, which can be understood by the fringing field effect. Semiconducting characteristics are successfully obtained by narrowing of the graphene channel, suggesting possible formation of a graphene nanoribbon under mild conditions, *i.e.*, in air at room temperature.

## Introduction

Two-dimensional (2D) atomic sheets, which are an elemental component of layered materials,[1,2] are ultimately thin materials. Thus, 2D sheets are inherently very sensitive to surface-related phenomena. Chemical modification of only the surface of 2D sheets results in alternation of the electronic structure of the whole body.[3-12]

The ultimately thin nature of 2D sheets enables us to tune the whole body by means of a field-effect transistor (FET) configuration. In the case of conventional thick semiconductors, the gate electric field is terminated near the interface with the gate dielectric, so the whole thickness cannot be controlled. On the other hand, in the case of 2D sheets, the gate controllability is perfect, *i.e.*, the whole thickness can be controlled because of the ultrathin nature. Therefore, the controllability of the FET configuration reaches the surface of ultrathin 2D sheets.

The surface controllability using an FET configuration raises the expectation of being able to control surface chemical modifications of 2D sheets. The FET-controlled chemical modification of surfaces has been reported on graphene, which is an archetypal 2D atomic sheet. A liquid-based functionalization of graphene with nitro-phenyl groups using aryl-diazonium salts was enhanced significantly by employing an electrochemical method (with a configuration of the so-called electric-double-layer transistor).[13] The occurrence of photo-oxidation of graphene by ultraviolet (UV) irradiation in ambient air was shown to be "switched off" by the polarity of the gate voltage.[14] The degree of phenyl-group functionalization of graphene surfaces using benzoyl peroxide was reported to be modified by changing the gate voltage.[15] These studies have proved that the FET configuration is indeed capable of controlling chemical modifications of the surface of 2D sheets.

Among the FET-controlled modifications described above, photo-oxidation of graphene affects the electronic structure of the 2D sheet itself to the highest degree. In the case of other reactions of grafting functional groups to graphene,[13,15] characteristic Raman features of graphene were still discernible after the reactions. On the other hand, the Raman scattering spectra of the photo-oxidized part after the FET-controlled reaction[14] showed a loss of characteristic features of graphene. In addition, photo-oxidation was found to initiate at the edges of the graphene channel and proceed towards the channel center. This indicates that the width of the graphene channel can be controlled by means of the FET-controlled photo-oxidation.

However, little is known about the detailed mechanism of the FET-controlled photo-oxidation. Counterintuitively, while oxidation is generally enhanced under electron accumulation because oxidation is a reaction defined as the loss of electrons, hole accumulation to graphene by applying a negative gate voltage was found to enhance FET-controlled photo-oxidation.[14] In addition, the FET-controlled photo-oxidation can narrow the graphene channel under mild conditions, *i.e.*, in air at room temperature. However, the FET-controlled oxidative narrowing has not been tested at a nanometer scale; in other words, graphene nanoribbon fabrication has not yet been attempted by this methodology.

In this study, the dependence of the degree of the reaction on various experimental parameters was investigated in order to discuss a possible mechanism for FET-controlled photo-oxidation of graphene. Among the parameters, the humidity of the experimental environment significantly affected the degree of oxidation; this can be related to the counterintuitive gate-polarity dependence owing to an adsorption-limited nature with a water-oxygen coadsorption picture. By optimizing the experimental parameters to enhance the degree of oxidation, the from-edge-to-center photo-oxidation was successfully exploited to narrow down the graphene channel. As a result of narrowing, the on-off ratio of the drain current of the graphene FET was increased from ca. 3 to 50 at room temperature, which suggests that a finite bandgap was introduced by the formation of a graphene nanoribbon. The present method is capable of fine-tuning the bandgap/width of the nanoribbon by monitoring the drain current during the oxidative narrowing.

## Results and discussion

### Experimental parameters

[a.] *Department of Physics and Electronics, and Nanoscience and Nanotechnology Research Center, Osaka Prefecture University, Sakai 599-8570, Japan. E-mail: r-nouchi@pe.osakafu-u.ac.jp*
[b.] *PRESTO, Japan Science Technology Agency, Kawaguchi 332-0012, Japan*
† Electronic Supplementary Information (ESI) available: Spectral radiation irradiance of a UV light used in this study.

Fig. 1 depicts details of the experimental procedure with various experimental parameters listed. Fig. 1a shows an optical micrograph of a fabricated FET with a mechanically exfoliated graphene flake as the channel material, which was deposited onto a highly doped Si wafer with a 300-nm-thick thermal oxide layer on top of it. As shown schematically in Fig. 1b, the fabricated FET was irradiated by a UV light through an optical fiber (details of the light source will be described in the Experimental section). Fig. 1c shows the time evolution of the UV irradiation procedure. First, the gate and drain voltages are applied, followed by UV irradiation after a certain waiting period. The experimental parameters include FET-related parameters, such as the drain/gate voltage and the channel shape, environmental parameters, such as the temperature and humidity of the ambience, and the procedure-related parameter of waiting time before starting the UV irradiation.

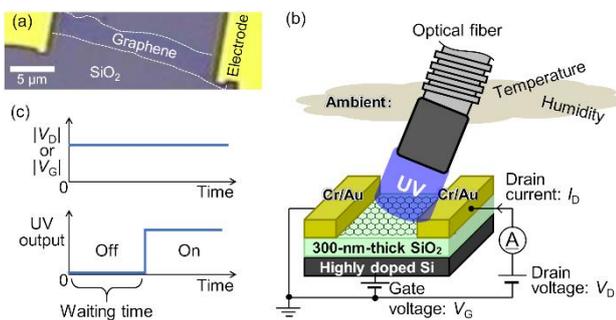

**Fig. 1** Device configuration and experimental parameters for gate-controlled photo-oxidation of graphene. (a) Optical micrograph of a fabricated graphene FET. Experimental parameters related to (b) FET operation and the surrounding environment and (c) UV irradiation procedure.

### Humidity-enhanced nature

First, we consider the effect of the humidity of ambient air. Fig. 2a and Fig. 2b show the Raman mapping of two graphene FETs after 10 min of UV irradiation in ambient air with different humidity values. The mappings are shown as a false color image, where the $D$ (magenta) and $2D$ (cyan) mappings are overlaid. The electrodes are also colored in magenta because of a background signal from the Au film. The gate and drain voltages during the UV irradiation were set to −60 V and 10 mV, respectively; the waiting time before starting the irradiation was ca. 1 min. The intensity of the Raman $D$ band is indicative of the degree of oxidation.[16] As is seen clearly from these figures, the UV irradiation under an absolute humidity of 9.7 g m$^{-3}$ caused photo-oxidation around the edges of the graphene flakes (Fig. 2a), while no photo-oxidation was observed after irradiation under an absolute humidity of 4.3 g m$^{-3}$ (Fig. 2b).

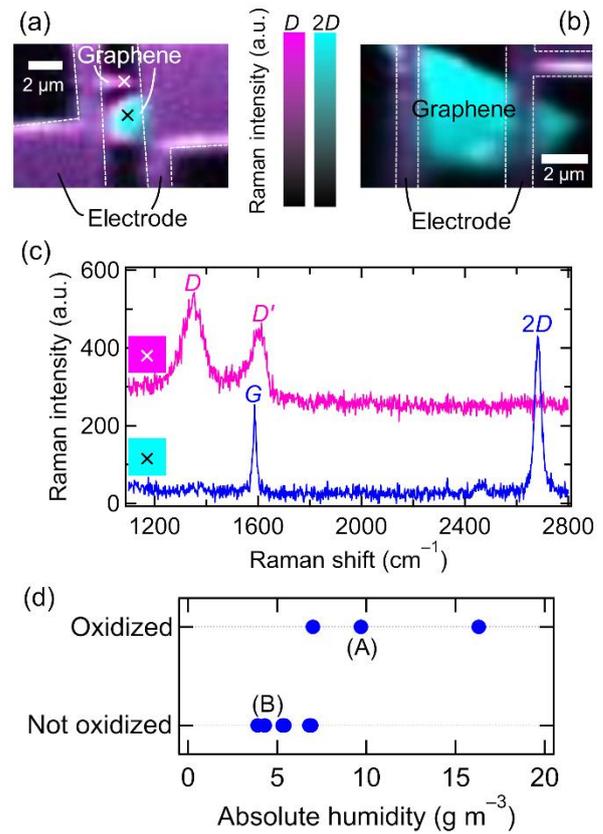

**Fig. 2** Effect of humidity on gate-controlled photo-oxidation of graphene. Raman mappings after UV irradiation at room temperature under (a) high and (b) low humidity. Other parameters were set as follows: the drain voltage was 10 mV, the gate voltage was −60 V, and the waiting time before starting the UV irradiation was ca. 1 min. (c) Raman scattering spectra taken at the oxidized (white cross) and non-oxidized (black cross) points in (a). (d) Compilation of the data acquired in uncontrolled ambience against the absolute humidity. Absolute humidities higher than ~7 g m$^{-3}$ resulted in photo-oxidation of graphene.

The Raman spectra taken at the oxidized and non-oxidized points in Fig. 2a are shown in Fig. 2c. At the oxidized point near the edges, a characteristic feature of graphene in the Raman spectrum (the $2D$ band) almost disappeared after UV irradiation; instead, the $D$ and $D'$ bands appeared. On the contrary, at the non-oxidized point near the channel center, the characteristic features survived even after UV irradiation, and the $D$ band was almost undetectable. These results show that the FET-controlled photo-oxidation of graphene is edge-selective and exerts almost no damage on the channel center.

Fig. 2d compiles the data acquired under an uncontrolled ambience with respect to the absolute humidity of the experimental environment. The environmental parameters (temperature and relative humidity) determine the value of the absolute humidity. Although the other parameters (i.e., the channel shape) should affect the result to some extent, as will be discussed later, the compiled data set shows a clear trend that the FET-controlled photo-oxidation of graphene is enhanced by increasing the humidity, as has been observed in photo-induced transformation of $WS_2$ in the presence of ambient moisture[17] and photo-oxidation of graphene in the presence of an intentionally formed water layer on the supporting substrate.[18]

**Peculiar dependence on gate-voltage polarity**

Although oxidation is generally enhanced under electron accumulation because oxidation is a reaction defined as the loss of electrons, FET-controlled photo-oxidation is enhanced under hole accumulation by applying a negative gate voltage.[14] Such peculiar dependence on the gate-voltage polarity is shown in the Raman mappings of Fig. 3a and Fig. 3b, where UV light irradiation was performed on the film for 20 min under the positive (+60 V) and negative (−60 V) gate voltages, respectively. The absolute humidity was approximately 12 g m$^{−3}$, the drain voltage was set to 1 mV, and there was a waiting time of 3 min before starting the UV irradiation. As shown in Fig. 3a and Fig. 3b, the photo-oxidation was enhanced when graphene was hole-doped using negative gating during the UV irradiation, which was further confirmed with three other device sets.

A possible mechanism accountable for the counterintuitive dependence on the gate voltage is proposed based on the fact that photo-oxidation is enhanced by humidity. In the humidity-enhanced photo-oxidation process, two molecules should be involved, namely $O_2$ to produce reactive oxygen species and $H_2O$ to contribute to the humidity-enhanced nature. First, it is assumed that the adsorption kinetics of $H_2O$ onto graphene is faster than that of $O_2$. This should be a reasonable assumption because $H_2O$ molecules possess a permanent electric dipole, which means they can adsorb more strongly onto graphene through dipole–induced dipole interaction. Thus, the first monolayer on graphene is considered to be a $H_2O$ layer, as shown in Fig. 3c and Fig. 3d. Then, $O_2$ molecules adsorb onto the $H_2O$ layer as an over-layer; a study based on molecular dynamics simulation[19] has shown that $H_2O$ molecules form a double-layer structure on graphene, but here the second $H_2O$ monolayer is omitted for simplicity. When the gate voltage is applied to a graphene FET, electric field lines are formed, as schematically depicted in Fig. 3c and Fig. 3d. On the outer surface of the graphene FET, almost no electric field is generated around the center of the graphene channel. However, a finite electric field, called a fringing field, is generated near the edges.[20, 21] For the present device with the channel width of ~2 μm and the gate dielectric thickness of 300 nm, the fringing field strength is several times higher than the corresponding wide-channel-limit value of the gate electric field (60 V/300 nm).[22] On the other hand, a density functional theory calculation has shown that an external electric field of ~1 V/nm is required to flip the dipole direction of $H_2O$ adsorbed on Au(111) and Au(110) surfaces,[23] which is close to the fringing field strength of the present study. As a result of the finite electric field strength of the outer surface around the edges, the orientation of $H_2O$ molecules becomes aligned with the field line because of the electric dipole of $H_2O$. Thus, the gate voltages with different polarities align $H_2O$ molecules in different directions.

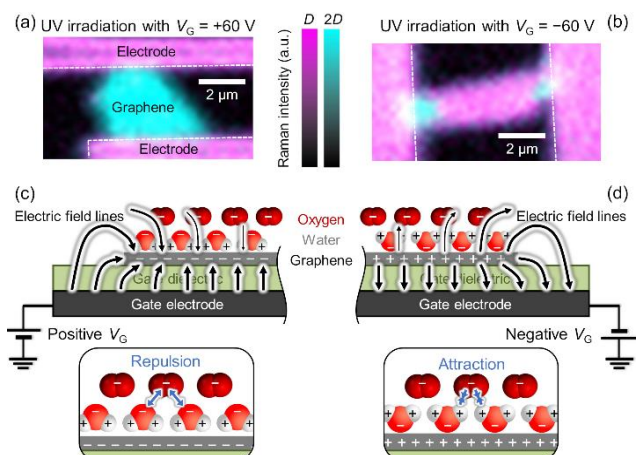

**Fig. 3** Effect of gate voltage polarity on the gate-controlled photo-oxidation of graphene. Raman mappings after UV irradiation at room temperature for (a) positive (+60 V) and (b) negative (−60 V) gate voltage. Other parameters were the drain voltage of 1 mV, waiting time of 3 min, and absolute humidity of ca. 12 g m$^{−3}$. Effect of a fringing field on a water-oxygen co-adsorption picture for (c) positive and (d) negative gating. The fringing electric field for positive (negative) gate voltage decreases (increases) the adsorption energy of oxygen molecules owing to the repulsive (attractive) forces from the water molecules oriented due to the fringing gate electric field.

The gate-controlled orientation of $H_2O$ molecules should be the key to understanding the peculiar gate-polarity effect. When the gate voltage is negative, $H_2O$ molecules adsorbed near the edges align such that their dipoles point away from the graphene surface, *i.e.*, the negatively charged oxygen side faces the graphene surface. When the gate voltage is positive, the positively charged hydrogen side faces the graphene. On the other hand, $O_2$ molecules on the $H_2O$ layer should be negatively charged, which is due to electron transfer from the underlying graphene. The electron transfer is known to be mediated by an oxygen/water redox couple through the following reaction:

$$O_2(aq) + 4H^+ + 4e^- (graphene) \underset{red}{\overset{ox}{\rightleftarrows}} 2H_2O \qquad (1)$$

This occurs under slightly acidic condition, where water molecules supply solvated oxygen molecules.[24, 25] The electron transfer from graphene to the redox couple was confirmed by measurements of transfer characteristics of graphene FETs in various gas atmospheres.[25] Compared to the characteristics measured in a vacuum, the transfer characteristics measured in the atmosphere containing both $O_2$ and $H_2O$ were found to show an overall shift toward positive gate voltages, which indicates that the electron transfer from graphene occurred within the studied gate-voltage range (from -40 V to +40 V with 285-nm-thick $SiO_2$ as the gate dielectric).[25] This result supports the $O_2$ layer being negatively charged irrespective of the polarity of the gate voltage. The negatively charged $O_2$ molecules feel an attractive or repulsive force with the hydrogen or oxygen side of the $H_2O$ layer. Thus, as schematically shown in Fig. 3c and Fig. 3d, the negative or positive gate voltage respectively results in a higher or lower binding energy of $O_2$ molecules on the oriented $H_2O$ layer.

There are two competitive processes taking place upon UV irradiation in air, namely desorption of $O_2$ molecules[26, 27] and formation of reactive oxygen species mediated by UV-induced generation of $O_3$.[16, 28] If the former process is dominant, no oxidation of graphene occurs. Thus, it is necessary to suppress the former process in order to observe photo-oxidation of graphene. When graphene is negatively gated, the adsorption energy of the $O_2$ molecules becomes higher than that in the positively gated case, as discussed above. The negative gate voltage should reduce the desorption probability of $O_2$ molecules, resulting in enhanced photo-oxidation of graphene. As a result of the suppression of the former process, the latter process becomes dominant. The light source used in this study, whose spectrum is shown in the Electronic Supplementary Information (Fig. S1), contains a small amount of light with sufficiently short wavelengths to crack oxygen molecules. Therefore, ozone and excited oxygen atoms should subsequently be formed. In addition, the excited oxygen atoms interact with $H_2O$ molecules and generate hydroxyl radicals. These reactive oxygen species should form various types of oxygen-related groups on the graphene surface, such as C-O-C, C-OH, and C=O, as shown in studies on graphite oxide.[29]

**Electrode effect**

By carefully looking at the Raman mapping of Fig. 3b, it was found that the Raman *D* band near the source/drain electrodes is rather small. The same trend was also observed in our previous study.[14] As will be shown later, the photo-oxidation possesses a thermally activated nature, and the low reactivity near the electrodes might be attributed to heat dissipation to the electrodes.[14]

The electrode metals of Cr and Au used in this study can both act as a catalyst. It was shown that deposition of Au nanoparticles onto graphene kinetically enhances the laser-induced photo-oxidation of graphene.[18] This indicates that the Au nanoparticles acted as a catalyst, possibly by generation of an evanescent electric field on the graphene surface. In addition, Cr is known to act as a thermal catalyst that corrodes graphene.[30] The experimental observations (*i.e.*, the low reactivity near the electrodes) indicate that possible catalytic activity of the electrode metals is not significant under our experimental conditions.

**Edge selectivity**

The edges of graphene are known to possess higher chemical reactivity than the basal plane because many dangling bonds exist at the edges.[31] Thus, it is natural to attribute the edge selectivity in the gate-controlled photo-oxidation to the presence of dangling bonds at the edges. However, the in-plane honeycomb structure is not free from defects, and the presence of dangling bonds is also unavoidable in the basal plane. The defect density in the basal plane *before* oxidation is too small to be detected by Raman spectroscopy in highly crystalline graphene samples fabricated by mechanical exfoliation. However, large etch pits in the basal plane were reported to be formed *after* a thermal treatment in an $Ar/O_2$ mixture, which was shown to be detectable by Raman spectroscopy.[32, 33] An atomic force microscope image taken after the thermal oxidation (Fig. 1 of Ref. 33) suggests that diameters of the etch pits formed in the basal plane seem to be comparable to a decrease in width due to the etched part at the edge of the same flake. This fact suggests that there is no large difference between reactivities of the in-plane defects and the edges toward thermal oxidation. The same discussion may hold for our case also, though our oxidation conditions are different from those in Ref. 33. If chemically active dangling bonds determine the edge selectivity, oxidation should occur over the whole graphene surface due to widely distributing defects in the basal plane, leading to an observation of the Raman *D* band throughout the graphene flake. However, this is not the case in the gate-controlled photo-oxidation because no oxidized patch was detected in the Raman mapping analyses. Therefore, the edge selectivity of the gate-controlled photo-oxidation is considered to be caused by the fringing electric field around the graphene edges.

Another possible explanation of the edge selectivity is permeation of $H_2O$ molecules into the graphene/substrate interface only around the edges. In the case that $H_2O$ molecules are present beneath the graphene layer, the orientation of the electric dipole of the $H_2O$ molecules becomes opposite to that described in Figs. 3c and 3d because the direction of the gate electric field becomes opposite to that of the case with $H_2O$ adsorbed on the graphene flake. As was discussed above, the gate-controlled photo-oxidation of graphene occurred with application of negative gate voltages. The electric dipole of the $H_2O$ molecules beneath graphene should be oriented downward (*i.e.*, the negative pole faces the graphene) under application of the negative gate voltages. In this case, the Coulomb interaction between negatively charged $O_2$ molecules on graphene and the $H_2O$ molecules beneath graphene should be repulsive, though the Coulomb interaction is partly screened by the presence of graphene. Therefore, under the model with $H_2O$ molecules underneath graphene, application of negative gate voltages should weaken the adsorption of $O_2$ molecules, leading to $O_2$ desorption upon photo-irradiation. This situation is inconsistent with another experimental finding that $O_2$ desorption occurs with positive gate voltages.[14] Therefore, the edge selectivity found in our study is again attributable to the fringing field.

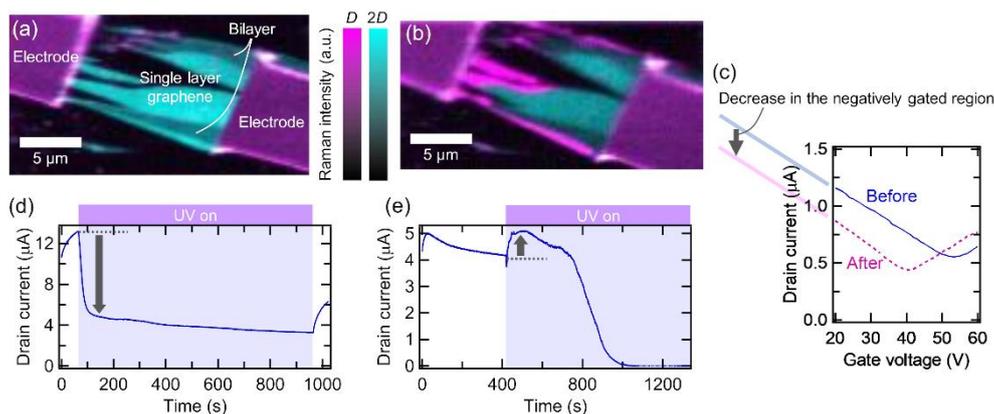

**Fig. 4** Effect of waiting time prior to UV irradiation on gate-controlled photo-oxidation of graphene. Raman mappings after UV irradiation with a waiting time of (a) 1 min and (b) 7 min. Other parameters were the drain voltage of 10 mV, gate voltage of −60 V, temperature of 27 °C, and absolute humidity of 15 g m$^{−3}$. (c) Transfer characteristics before and after the UV irradiation with a waiting time of 1 min. (d) and (e) are the time evolution graphs of the drain current during the UV irradiation for (a) and (b), respectively.

**Sufficient adsorption of water**

In order to guarantee a high binding energy of $O_2$ molecules, a sufficient number of $H_2O$ molecules should adsorb onto the graphene before irradiation of UV light. Otherwise, an insufficient number of adsorbed $H_2O$ molecules will result in UV-induced $O_2$ desorption. Charge carrier doping to graphene has been theoretically shown to lead to higher water adsorption,[34] which is supported by the experimental fact that water wettability is enhanced upon electrostatic doping of graphene by gating.[35, 36] Thus, prolonged application of a gate voltage, which electrostatically dopes graphene, is useful to guarantee the adsorption of a sufficient amount of $H_2O$ molecules.

Fig. 4a and Fig. 4b show the Raman intensity mappings acquired after UV irradiation with identical FET-related/environmental parameters, but with different waiting times before starting the UV irradiation. In this particular device, the short waiting time of 1 min resulted in no photo-oxidation (Fig. 4a). However, after UV irradiation to an identical device with a longer waiting time of 7 min, the $D$ band clearly appeared around the edges (Fig. 4b). The fact that the longer waiting time causes a higher degree of photo-oxidation is consistent with the possible adsorption of more $H_2O$ molecules by the prolonged gate-voltage application.

The results in Fig. 4a and Fig. 4b correspond to UV-induced $O_2$ desorption and photo-oxidation, respectively. The former result should be accompanied by dedoping of graphene because oxygen molecules are known to dope holes to graphene.[37] The hole dedoping was confirmed by the shift of the transfer characteristics after the UV irradiation with a waiting time of 1 min (the shift of the minimum conductivity point towards the negative gate voltage direction in Fig. 4c). The $O_2$ desorption is also featured in the time evolution of the drain current during the corresponding UV irradiation procedure (Fig. 4d). Hole dedoping upon $O_2$ desorption decreases the drain current under the negative gate-voltage application, as schematically depicted by the arrow in Fig. 4c, which was indeed observed in Fig. 4d (indicated by an arrow). On the other hand, when the $O_2$ desorption is suppressed, as it is with a waiting time of 7 min, UV irradiation should increase the drain current because of photoconductivity of graphene, which is indicated in Fig. 4e. Therefore, the series of the experimental results shown in Fig. 4 is consistent with the water-oxygen co-adsorption picture.

It should be noted here that the drain current increase is not always a sign of photo-oxidation. As shown in Fig. 3c and Fig. 3d, the fringing electric field is significant only near the edges of the channel. Thus, regardless of the polarity of the gate voltage, UV-induced $O_2$ desorption occurs near the channel center where the fringing gate field is weak. If the channel width is large, the drain current can decrease at the same time as photo-oxidation near the channel edges, which is due to the large contribution of $O_2$ desorption around the channel center.

**Thermally activated nature**

A device identical to that shown in Fig. 4b was subsequently used to examine the effect of temperature. The environmental parameters, *i.e.*, the temperature and humidity of the ambience, were controlled by a thermo-hygrostat. To avoid Joule heating, no drain voltage was applied to the device, and the source/drain terminals were grounded. Only a gate voltage of −60 V was applied to the device during the UV irradiation. Fig. 5 shows the Raman intensity mappings acquired after each UV irradiation period under different environmental conditions. The initial state (Fig. 5a) is identical to that in Fig. 4b, where oxidation-induced discontinuity of the channel near the left electrode further ensures no temperature rise by Joule heating.

From the series of Raman data in Fig. 5, the photo-oxidation by the UV light irradiation with application of no drain voltage was found to be suppressed when the temperature was down to 40 °C, even if the absolute humidity was set to a high value of 30 g m$^{−3}$ (Fig. 5d). The oxidation evolution speed was estimated using Figs. 5b, 5d, and 5e; its temperature dependence is plotted in Fig. 5f. The thermally activated nature of the photo-oxidation is clearly depicted.

In Fig. 5c, the relatively low absolute humidity (17 g m$^{−3}$) resulted in no oxidation even at the highest temperature tested of 50 °C. On the other hand, as shown in Fig. 2c, UV irradiation under application of a finite drain voltage was found to induce photo-oxidation with an absolute humidity down to approximately 7 g m$^{-3}$. Thus, the drain voltage/current is considered to act as an additional heat source that generates Joule heat, which enhances the thermally activated photo-oxidation. Indeed, a higher value of the drain voltage was found to result in a higher degree of photo-oxidation in our previous work.[14]

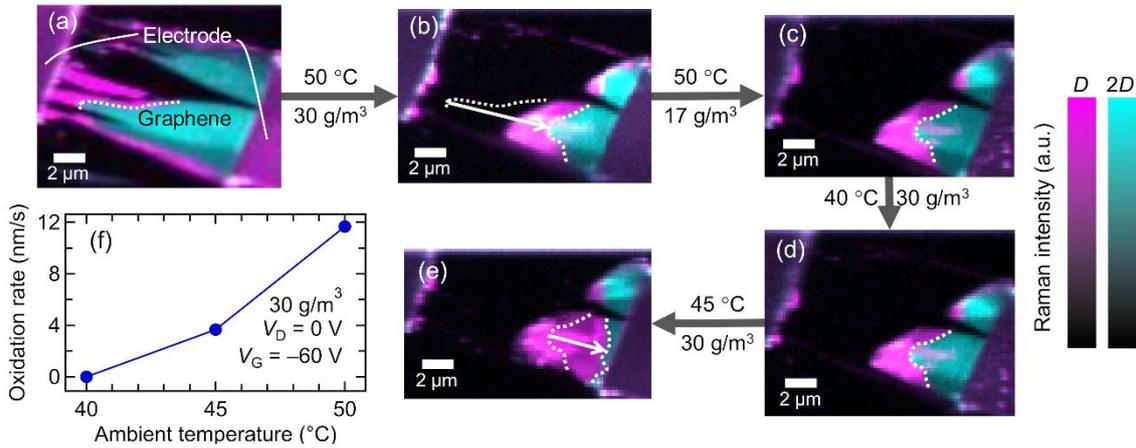

**Fig. 5** Effect of ambient temperature on gate-controlled photo-oxidation of graphene. Raman mappings were acquired sequentially after UV irradiation under different ambient conditions controlled by a thermo-hygrostat: (a) initial state identical to that in Fig. 4b at a temperature and absolute humidity of (b) 50 °C and 30 g/m$^3$, (c) 50 °C and 17 g/m$^3$, (d) 40 °C and 30 g/m$^3$, and (e) 45 °C and 30 g/m$^3$, respectively. In addition to the source terminal, the drain terminal was grounded. Other parameters were the gate voltage of −60 V and waiting time of 8 min. (f) Oxidation rate as a function of temperature. The data in (f) were calculated from the Raman mappings of (b), (d), and (e).

Joule heat causes temperature rise inside the graphene channel. However, the temperature rise is not uniform because of the nonuniformity of the channel width, which causes a nonuniform current density. Fig. 6 shows the Raman intensity mapping before and after UV irradiation to an FET of a graphene flake with a narrowed area. After the UV irradiation with a drain voltage of 20 mV, the narrowed area was selectively oxidized, as expected from the larger temperature rise caused by the higher current density of the area.

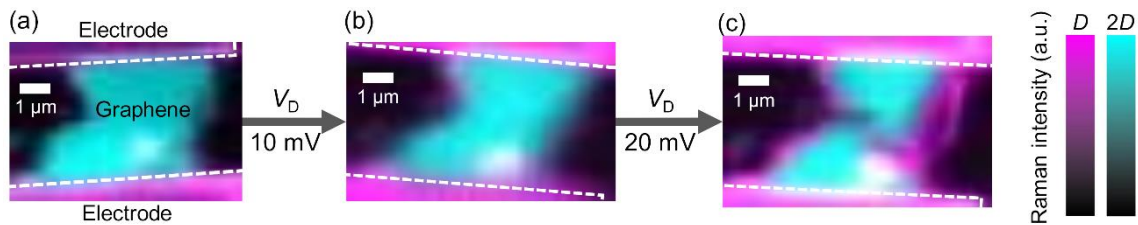

**Fig. 6** Effect of channel shape on gate-controlled photo-oxidation of graphene. Raman mappings acquired (a) at the initial state and after UV irradiation with the drain voltage of (b) 10 mV and (c) 20 mV. Other parameters were the gate voltage of −60 V, temperature of 27 °C, absolute humidity of 15 g m$^{-3}$, and waiting time of 1 min. The narrowed region was selectively oxidized in (c).

**Electronic structure control by narrowing**

The gate-controlled photo-oxidation of graphene was found to initiate at the edges and proceed into the channel center. This indicates narrowing of the conducting graphene channel. Graphene is known to possess a finite bandgap when its width is narrowed down to the nanometer scale.[38-40] As a result of narrowing, the conducting graphene channel is expected to become a semiconducting graphene nanoribbon.

Next, this from-edge-to-center photo-oxidation was exploited to fabricate a graphene nanoribbon. Fig. 7a shows the Raman mapping of a single-layer graphene flake used for this purpose. The edge of the flake shows a corrugation, and thus a slightly narrowed region exists in the channel. UV light was irradiated on this device under preferable experimental conditions described above in order to enhance photo-oxidation: negative gate voltage (−60 V), high humidity (15 g m$^{-3}$), high drain voltage (20 mV), and long waiting time before UV irradiation (7 min). After the UV irradiation, the slightly narrowed region was found to be predominantly oxidized as shown in Fig. 7b, which can be explained by the nonuniform Joule heat, as in Fig. 6. The two-terminal FET characteristics (transfer characteristics) were measured at room temperature, as shown in Fig. 7c. The current on-off ratio was found to increase from ca. 3 to 50, which suggests the formation of a semiconducting graphene nanoribbon by the gate-controlled oxidative narrowing.

The present method enables us to fabricate a semiconducting graphene nanoribbon under mild conditions, *i.e.*, in air and at room temperature. The fabrication methods of graphene nanoribbon reported to date require a fine lithographical patterning,[41-43] bottom-up chemical synthesis,[44-46] unzipping of carbon nanotubes,[47, 48] edge etching by heat and/or plasma treatment,[49-51] etc. Among them, only the present method potentially enables nanoribbon fabrication in air at room temperature.

In addition, the gate-controlled photo-oxidation method allows us to monitor the electric current during the oxidative narrowing process because this method exploits a graphene FET under operation (*i.e.*, with application of the drain and gate voltages). Thus, this method potentially enables fabrication of a semiconducting graphene channel with a controlled bandgap by carefully selecting the value of the electric current at which to stop the UV irradiation, which is the future prospect of this study.

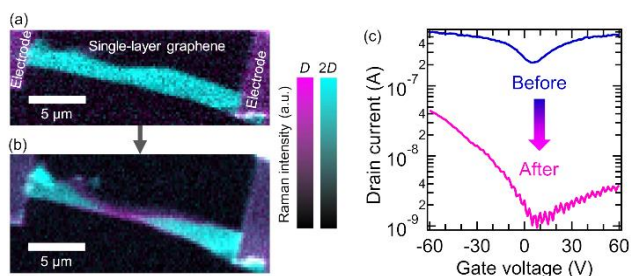

**Fig. 7** Semiconducting narrow graphene channel formed by gate-controlled photo-oxidation of graphene. Raman mappings (a) before and (b) after UV irradiation. Experimental parameters were a gate voltage of −60 V, drain voltage of 20 mV, waiting time of 7 min, temperature of 27 °C, and absolute humidity of 15 g m$^{-3}$. (c) Transfer characteristics of the graphene FET before and after the UV irradiation.

## Conclusions

Gate-controlled photo-oxidation of graphene was investigated in detail by considering effects of various experimental parameters. The reaction, which possesses a thermally activated nature, was found to be enhanced by water adsorption on graphene. The water-enhanced nature was exploited to explain the peculiar dependence on the gate-voltage polarity, where the gate-controlled photo-oxidation was enhanced by electrostatic hole doping under negative gate voltages although the electron-withdrawing reaction (*i.e.*, oxidation) should be enhanced by electron doping. The counterintuitive gate-voltage dependence is attributable to the gate-induced orientation of water molecules, which acts as the stabilizer of oxygen adsorption, as shown in Fig. 3d. In addition, the possible catalytic effect of the electrode metals was found to be negligable under our oxidation conditions. The graphene oxide formed by the gate-controlled photo-oxidation is considered to have a chemical structure of C-O-C, C-OH, and/or C=O, but to confirm this it is necessary to perform X-ray photoelectron microscopy analyses.

The gate-controlled photo-oxidation of graphene initiated at the edges of the graphene channel and proceeded towards the channel center, indicating that the conducting graphene channel can be narrowed by this method. By exploiting this method, we have succeeded to increase the current on-off ratio from ca. 3 to 50. This result suggests successful conversion of a wide conducting graphene layer to a narrow semiconducting graphene nanoribbon. The present method is expected to fine-tune the bandgap/width of the nanoribbon by monitoring the drain current during gate-controlled oxidative narrowing of graphene.

## Experimental

A highly doped Si wafer with a thermally grown 300-nm-thick oxide layer on top of it was used as a substrate for the FET fabrication. The doped Si layer acts as the gate electrode while the SiO$_2$ layer serves as the gate dielectric. The substrate was cleaned with acetone and isopropanol in an ultrasonic bath followed by oxygen plasma treatment (Harrick Plasma, PDC-32G). Graphene flakes were formed by mechanical exfoliation[52] onto the cleaned substrate. Au electrodes with a thin Cr adhesion layer (1 nm) as the source/drain electrodes were fabricated by conventional electron-beam lithography processes.

For gate-controlled photo-oxidation, a source-measure unit (Keithley, 2636A; or Keysight, B2902A) and a UV irradiation system (Ushio, SP7-250UB; the spectral radiation irradiance is shown in the Electronic Supplementary Information, Fig. S1) equipped with a deep UV lamp (Ushio, UXM-Q256BY) were used. The irradiance at the sample was ca. 1.6 W cm$^{-2}$ for the data in Figs. 2 and 3 and 2.6 W cm$^{-2}$ for the data in Figs. 4–7. The spot size of the UV light was sufficiently larger than the device size, which ensures irradiation of the whole graphene channel. A thermo-hygrostat (Eyela, KCL-2000A) was used for the experiments where the ambient temperature and humidity must be controlled. The degree of oxidation was determined using a Raman microscope (Nanophoton, Raman DM).

## Conflicts of interest

There are no conflicts to declare.

## Acknowledgements

This work was supported in part by the Special Coordination Funds for Promoting Science and Technology from the Ministry of Education, Culture, Sports, Science and Technology of Japan; and JSPS KAKENHI Grant Numbers JP26107531, JP16H00921, and JP17H01040.